\documentclass{article}

\usepackage{microtype}
\usepackage{graphicx}
\usepackage{subfigure}
\usepackage{subcaption}
\usepackage{booktabs} 

\usepackage{hyperref}

\usepackage[accepted]{icml2024}

\usepackage{amsmath}
\usepackage{amssymb}
\usepackage{mathtools}
\usepackage{amsthm}
\usepackage{multirow}
\usepackage{algpseudocode}
\usepackage{comment}
\usepackage[normalem]{ulem}

\usepackage[capitalize,noabbrev]{cleveref}

\theoremstyle{plain}

\theoremstyle{definition}

\theoremstyle{remark}

\usepackage[textsize=tiny]{todonotes}

\icmltitlerunning{Generative Model for Small Molecules with Latent Space RL Fine-Tuning to Protein Targets}

\begin{document}

\twocolumn[
\icmltitle{Generative Model for Small Molecules with Latent Space \\ RL Fine-Tuning to Protein Targets}



\icmlsetsymbol{equal}{*}

\begin{icmlauthorlist}
\icmlauthor{Ulrich A. Mbou Sob}{equal,comp}
\icmlauthor{Qiulin Li}{equal,comp,sch}
\icmlauthor{Miguel Arbesú}{comp}
\icmlauthor{Oliver Bent}{comp}
\icmlauthor{Andries P. Smit}{comp}
\icmlauthor{Arnu Pretorius}{comp}
\end{icmlauthorlist}

\icmlaffiliation{comp}{InstaDeep}
\icmlaffiliation{sch}{Work done during internship at InstaDeep}

\icmlcorrespondingauthor{Ulrich A. Mbou Sob}{u.mbousob@instadeep.com}

\icmlkeywords{Drug discovery, Reinforcement Learning, Large Language Models}

\vskip 0.3in
]



\printAffiliationsAndNotice{\icmlEqualContribution} 

\begin{abstract}
A specific challenge with deep learning approaches for molecule generation is generating both syntactically valid and chemically plausible molecular string representations. To address this, we propose a novel generative latent-variable transformer model for small molecules that leverages a recently proposed molecular string representation called SAFE. We introduce a modification to SAFE to reduce the number of invalid fragmented molecules generated during training and use this to train our model. Our experiments show that our model can generate novel molecules with a validity rate $>$ 90\% and a fragmentation rate $<$ 1\% by sampling from a latent space. By fine-tuning the model using reinforcement learning to improve molecular docking, we significantly increase the number of hit candidates for five specific protein targets compared to the pre-trained model, nearly doubling this number for certain targets. Additionally, our top 5\% mean docking scores are comparable to the current state-of-the-art (SOTA), and we marginally outperform SOTA on three of the five targets.
\end{abstract}

\section{Introduction}
\label{sec:intro}
De novo drug design is the process of designing novel chemicals with desired pharmacological properties. This process is time-consuming and requires screening numerous compounds to find potential drug candidates \citep{bohacek1996art}. The chemical space for potential drugs is vast, up to $\sim 10^{60}$ \citep{polishchuk2013estimation}, and sparse, making exploration challenging. Deep learning methods for drug design aim to optimize the search process and reduce the time required for drug discovery and development (see \citet{atance2022novo} and \citet{lee2023exploring}).  

Deep learning approaches for molecule generation can be categorized into three groups: \emph{sequence} based, \emph{molecular graphs}, and \emph{structure} based approaches \citep{du2022molgensurvey}. Molecular graphs and structure-based approaches use graph and equivariant \citep{satorras2021n} neural networks, generally following a lead-based generation strategy, where new compounds are created by expanding or altering an existing core molecular fragment of known desirable properties with smaller functional groups --- i.e.\ sampling the chemical space around a scaffold. Examples of this approach include \citet{madhawa2019graphnvp}, \citet{zang2020moflow}, \citet{luo2021graphdf}, \citet{bongini2021molecular}, \citet{jiang2021could}, and \citet{powers2022fragment}. On the other hand, sequence-based approaches utilize molecular string representations such as SMILES \citep{smiles} and rely on sequence-based learning models like recurrent neural networks and transformers (see \citet{olivecrona2017molecular}, \citet{podda2020deep}, and \citet{irwin2022chemformer}). Lead-based generation strategies used in molecular and structure-based models are limited in their ability to generate molecules with completely new scaffolds and thus in diversity of the generated molecules. Sequence-based models, on the other hand, struggle to generate both syntactically valid representations and chemically plausible molecules with desired drug-like properties. For instance, deep learning models trained with SMILES often generate a high rate of syntactically invalid molecules, while models trained with SELFIES \citep{krenn2020self} tend to produce implausible molecules that typically consist of very long chains and large rings (see \citet{tarasov2023offline} and \citet{noutahi2024gotta}).
 
In this paper, we introduce a new generative latent-variable model for small molecules. This model consists of a variational auto-encoder (VAE) embedded within an encoder-decoder transformer architecture. We train our model using a novel molecular representation called \textbf{SAFER} which is based on the recently introduced \textbf{SAFE} representation \citep{noutahi2024gotta}. Additionally, we perform various experiments with this new architecture and apply reinforcement learning (RL) fine-tuning to generate potential hit candidates with high molecular docking scores to the same targets presented in \citet{lee2023exploring}.


\begin{figure}
    \centering
    \includegraphics[width=1\linewidth]{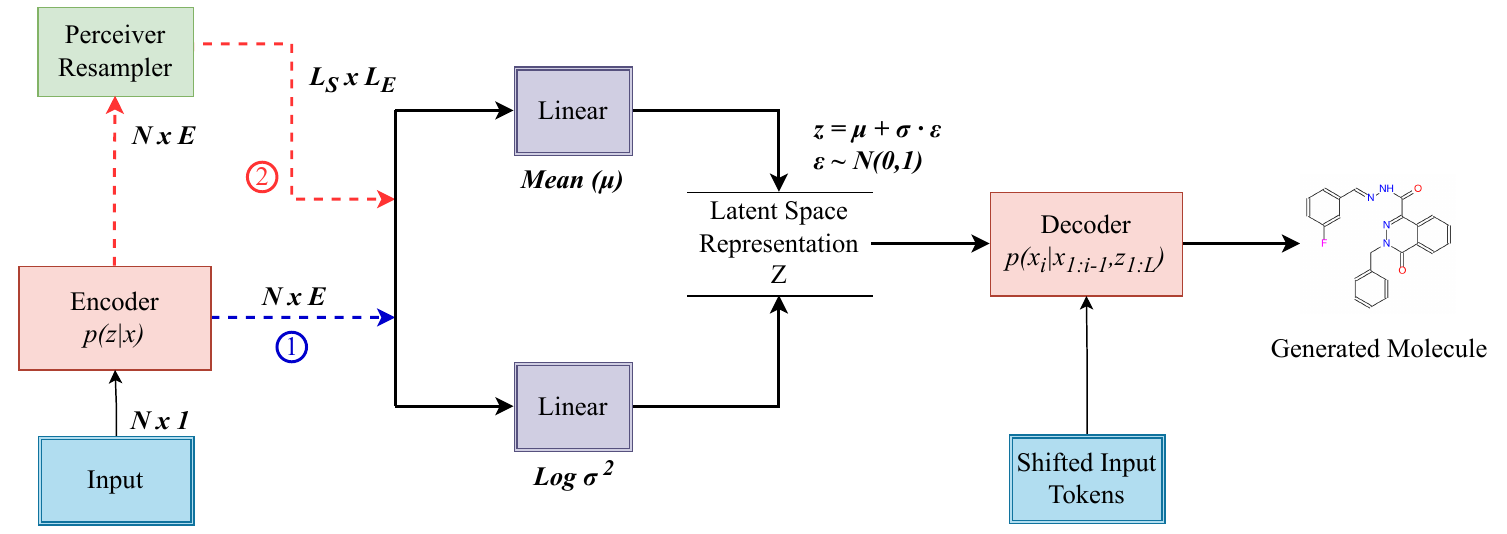}
    \caption{\textit{Schematic representation of our model's architecture.} A sequence of $N$ tokens is passed as input to our encoder which is a transformer model. The output encoded embeddings of shape $N\times E$ are either passed directly to the \texttt{mean} and \texttt{logvar} layers (path 1) or they are first passed to the perceiver resampler layer which maps the encoded embeddings to a reduced dimension of shape $L_S\times L_E$ (path 2). The \texttt{mean} and \texttt{logvar} layers are linear layers that are applied independently to each sequence dimension. The final reparametrised embeddings are then passed to the decoder transformer model to be used as encoder embeddings in the decoder's cross-attention layers.}
    \label{fig:VAE architecture}
\end{figure}

\section{Related Work}
\textbf{Molecular string representations.}\label{symmetries_review} The most popular and versatile notation for string encoding of molecules is the SMILES notation \citep{smiles}. Approaches such as \citet{olivecrona2017molecular}, \citet{gupta2018generative}, and \citet{gomez2018automatic} have trained different models such as RNNs for molecular generation using SMILES. Due to the difficulty of generating a high percentage of valid molecules with SMILES notation, various other notations have been proposed and different models have been trained using DEEPSMILES \citep{o2018deepsmiles} and SELFIES \citep{krenn2020self, feng2024gcardti}. Recently, the SAFE \citep{podda2020deep} representation was introduced, which in our opinion has the most potential to alleviate current challenges with single token string representations. A new study \cite{skinnider2024invalid} demonstrated that models that generate invalid molecules are more powerful than generate only valid molecules. However this study only compares SMILES and SELFIES. 

\textbf{Score-based approaches.} A major challenge in \textit{de novo} drug design is the need to solve a multi-objective problem, where the generated molecules need to possess various and usually unrelated physicochemical and pharmacological properties. Additionally, some of these properties are difficult to measure or computationally expensive to approximate with simulation tools such as AutoDock \citep{trott2010autodock} --- especially those related to the binding energy and affinity to a given target. Furthermore, there are a limited number of available datasets to train deep learning models to accurately predict these properties. Some attempts to learn molecular docking scores were made by \citet{corso2022diffdock} and \citet{ganea2021independent}. Different score-based approaches, such as using RL or diffusion models, have been proposed by \citet{olivecrona2017molecular}, \citet{jeon2020autonomous}, \citet{jo2022score}, \citet{lee2023exploring}, and \citet{tarasov2023offline}. Models such as \cite{lee2023exploring} have achieved SOTA performance in molecular docking. Additionally, multi-objective Bayesian methods, as shown by \cite{FROMER2023100678}, have demonstrated promising results in small molecules design. These approaches have mainly been applied to fine-tune models for specific single protein targets. The architecture proposed in this work opens the possibility of having a single model that can generalize to multiple protein targets if the structural intimation about the protein is provided to the model. 

\section{Methods}
\paragraph{Tokenization}\label{sec:safer}
In SMILES notation, digits identify opening and closing ring atoms. This makes it difficult to train deep learning models since multiple SMILES can correspond to the same molecule. Furthermore, ordering does not matter in SMILES notation, making it more challenging to train sequence models that pay attention to the position of the tokens in the sequence.

The SAFE tokenization modifies the SMILES notation by representing the molecule as a sequence of connected fragments. The molecule is fragmented using the BRICS algorithm \citep{degen2008brics}. Within the individual fragments, the SAFE tokenization preserves the ordering of the tokens but the ordering of interconnected fragments is not retained. Instead, the fragments are sorted by their molecular weights in decreasing order. 

\textbf{Challenges with SAFE}  We believe training a deep learning model using SAFE tokenization remains challenging due to the following two reasons:
\begin{enumerate}
\item The fact that the ordering of the fragments is not preserved produces a similar challenge as when using SMILES due to the positional ambiguity and multiple SMILES corresponding to the same molecule.
\item The model still has to learn to use digits to represent the opening and closing ring atoms. Hence the model suffers from the same challenges faced when using SMILES in its ability to generate valid molecules. 
\end{enumerate}

\textbf{A SAFER tokenizer for small molecules}  To improve upon SAFE tokenization, we propose the following modifications to the SMILES to SAFE conversion algorithm:
\begin{enumerate}
\item We canonicalize the input SMILES to enforce that each molecule has a unique SMILES notation.
\item After breaking the molecules into different fragments, we do not sort these fragments by their molecular weight. Instead, we preserve the ordering of the fragments as they appear in the original canonicalized sequence. 
\item We introduce two new tokens $[@]$ and $[@@]$ to denote open and closing ring atoms. Hence, instead of using digits to represent the ring atoms for the bonds that link the various fragments we instead use these two unique tokens. The need to preserve the ordering of the fragments arises as a consequence of the order in which the various attachment points must appear. This also reduces the necessity for the language model to learn different digit tokens which intrinsically do not have unique meanings. 
\end{enumerate}
Algorithm \ref{smiles_to_safe} shows the modifications we make to the SAFE algorithm to produce our SAFER tokenizer. The lines in blue correspond to the new steps we introduce and the strikeout text are the steps from the original algorithm that we do not use. Our approach takes inspiration from \cite{podda2020deep} where each fragment is expected to have one or two bonding atoms. Fragments at either end of the sequence have one bonding atom and fragments in the middle have two. It is important to note that our algorithm can fail if we have a fragment with more than two bonding atoms. However such cases are extremely rare and will only occur if we have fragments consisting of a single atom with valence $>$2. Fortunately, this can be prevented by constraining the bonds-breaking algorithm \citep{degen2008brics} similarly to how it is implemented in \cite{podda2020deep} to generate avoid breaking bonds that will lead to fragments such single atom fragments with more than two bonds. Table \ref{tab:smiles_example} provides an example molecule represented using the SMILES, SAFE and SAFER notations.

\begin{table}
\caption{Example of SMILES, SAFE and our SAFER representation for the same molecular string.}
\centering
\resizebox{\columnwidth}{!}{%
\begin{tabular}{|c|c|} 
\hline
Representation & String \\ 
\hline 
SMILES & CC(C)(C)c1ccc2occ(CC(=O)Nc3ccccc3F)c2c1 \\ 
SAFE & CC7(C)C.c17ccc2occ6c2c1.C6C4=O.N45.c15ccccc1F \\ 
SAFER & CC[@](C)C.c1[@@]ccc2occ[@]c2c1.C[@@]C[@]=O.N[@@][@].c1[@@]ccccc1F\\ 
\hline 
\end{tabular}
}
\vspace{10pt}
\label{tab:smiles_example}
\end{table}

\paragraph{Model Architecture}
Our objective is to train a powerful generative model for small molecules. Furthermore, with a latent-variable model design our generative model can perform \textit{de novo} molecule generation with specific properties by exploring different regions of the latent space distribution. Thus our goal is to learn a multi-dimensional latent distribution over the space of small molecules. The flowchart in Figure \ref{fig:VAE architecture} depicts our model's architecture. We leverage the success of transformers \citep{vaswani2017attention} in learning embeddings for various sequence tokens when trained in a self-supervised setting. We construct our architecture by adding a variational auto-encoder (VAE) latent space sampling layer in between the encoder and the decoder of the canonical transformer architecture. This is similar to the approach use in \citet{fang2021transformer} for controllable story generation.  

The input molecule is tokenized and passed to the encoder. The encoder, which is an N-Layer transformer model, computes embeddings for the input molecule. These encoded embeddings are then passed to a \texttt{mean} and \texttt{logvar} layer which computes the mean and log variance of the molecule's latent space distribution. Using the mean and log variance, we apply the reparameterization trick to sample the latent vectors which then serve as our final encoder embeddings to be passed to the decoder. The decoder, which is also a transformer, is conditioned to predict the next token in the sequence given the previous tokens. In order to scale our architecture while still maintaining the same latent space dimension, we insert an optional perceiver layer \citep{jaegle2021perceiver} which maps the encoded embeddings to a fixed sequence length and embedding dimension before passing the output to the \texttt{mean} and \texttt{logvar} layers. The perceiver achieves this by perfoming cross-attentions with learned query vectors that have the desired latent space dimensions.

\paragraph{RL Fine-Tuning to Protein Targets}
One of the most challenging aspects of \textit{de novo} drug design is to generate molecules with high binding affinity to specific target proteins. This binding affinity is usually predicted through an expensive computational process called molecular docking. Molecular docking typically consist of physics-based simulations used to find the most stable `docked' conformation of the generated molecule (ligand) with a specific binding site or pocket on target protein. Hence, a crucial step in the \textit{in silico} drug screening process is predicting the binding affinity of the candidate molecules to specific target sites. In this section, we demonstrate a proof-of-concept, for fine-tuning various parts of our architecture to generate molecules with high docking scores for a specific protein target.

We formulate our RL fine-tuning problem as follows: given an input molecule, encode and generate a new molecule from its embeddings, and provide a reward signal as the improvement in docking score over that of the original molecule. Thus the reward is defined as follows:
\begin{equation}
    R = \mathrm{DS_{new\_molecule}} - \mathrm{DS_{original\_molecule}} \label{reward}
\end{equation}
where DS denotes the docking score. The \texttt{mean} and \texttt{logvar} layers of our model are considered as policy parameters.

The model is fine-tuned using the REINFORCE algorithm \citep{williams1992simple} minimising the following loss function:
\begin{equation}
    \mathcal{L} = -\log \pi(a | s) \times R
\label{rl_loss},
\end{equation}
where $\pi(a|s)$ denotes our policy. The state $s$ is represented by the encoded embeddings of the original molecule whose docking score we aim to improve, while the action $a$ involves choosing the \texttt{mean} and \texttt{variance} of the region in our latent space from which to sample the modified molecule.

\section{Experiments}
In this section, we compare the performance of both our models and the SAFER representation. Recall that our main goal is to train a generative model for small drug-like molecules. To comprehensively evaluate the performance of our various models we measure the properties of the types of molecules generated. Hence, during evaluation we randomly sample a set of molecules using each model and compute the following quantitative metrics: Validity rate, Fragmentation rate, Uniqueness, Similarity,  Quantitative Estimate of Drug-likeness (QED) and Synthetic Accessibility (SA) (see Appendix \ref{sec:eval_metrics}).
To compare models during training we use the following combined metric:
\begin{equation}
    \begin{split}
        \mathrm{Validation\_Metric} = \mathrm{Validity\_Rate} \, \times  \\ (1 - \mathrm{Fragmentation\_Rate}) \times \mathrm{Mean\_QED} \label{sl_metric}
    \end{split}
\end{equation}
For drug-like molecules, we want our models to have a high validity rate, low fragmentation rate and high QED values.

\paragraph{Model Scaling}\label{sec:model_scaling}
We perform various experiments to evaluate how our model's performance scales with model size. Due to the need to have an architecture with a fixed latent dimension, we include a perceiver-resampler layer in these experiments to map the encoder's output to a fixed embedding dimension before passing these to the \texttt{mean} and \texttt{logvar} layers. We trained two sets of models: one with an embedding dimension of 128 ($\sim$4.5M parameters) and another with an embedding dimension of 256 ($\sim$16M parameters). For each embedding dimension, we trained another model in which we included the perceiver layer to map the embeddings to a smaller latent dimension equal to half of the original embedding dimension.

Table \ref{tab:scale-results} presents the average metrics from 10k samples generated, with the best parameters for each model using greedy decoding. All of our models are capable of generating molecules with a validity rate $>$ 90\% and a fragmentation rate $<$ 1\%. Additionally, all our models have a mean QED $>$ 0.75, an encouraging result. Interestingly, the model with an embedding dimension of 128 and the inclusion of the perceiver layer (Emb-128 + P-64) significantly outperforms all other models in terms of uniqueness (diversity). However, in terms of our selected validation metric, the model with an embedding dimension of 128 and without the perceiver (Emb-128) performs the best. The notable difference in uniqueness when the perceiver is applied to the model with a dimension of 64 (Emb-128 + P-64), suggests that an embedding dimension less than 128 is sufficient to represent the latent distribution of our dataset. In contrast, the results from the model with an embedding dimension of 256 are more challenging to interpret, as there are no noticeable differences between the models with and without the perceiver. 
\begin{table}
    \centering
    \caption{Average scores of 10k molecules sampled using different model configurations trained with our SAFER representation and temperature parameter set to 0. $\downarrow$ and $\uparrow$ indicate that the metric must be minimized or maximized, respectively. All metrics have been scaled to be in the range [0,1]. The first number on the configuration label indicates the embedding dimension and the number after ``+'' indicates the perceiver's output dimension for configurations in which we include the perceiver layer.}
    \label{tab:scale-results}
    \resizebox{\columnwidth}{!}{%
    \begin{tabular}{l|r|r|r|r|r|r}
        \toprule
        \textbf{Configuration} & \textbf{Validity} $\uparrow$ & \textbf{Frag rate} $\downarrow$ & \textbf{Uniqueness} $\uparrow$ & \textbf{QED} $\uparrow$ & \textbf{SA} $\downarrow$ & \textbf{Val Metric} $\uparrow$ \\
        \midrule
        Emb-128 & \textbf{0.97} & 0.002 & 0.59 & 0.86 & \textbf{0.72} & \textbf{0.83} \\
        Emb-256 & 0.91 & 0.013 & 0.73 & \textbf{0.87} & 0.76 & 0.782 \\
        \midrule
        Emb-128 + P-64 & 0.90 & 0.0498 & \textbf{0.855} & 0.75 & 0.87 & 0.65 \\
        Emb-256 + P-128 & 0.92 & \textbf{0.0015} & 0.71 & 0.86 & 0.76 & 0.788 \\
        \bottomrule
    \end{tabular}%
    }
\end{table}
\paragraph{Model Finetuining}
We \textit{independently} fine-tune our model for the same five human target proteins in \cite{lee2023exploring} i.e.\ PARP-1 (Poly [ADP-ribose] polymerase 1 ), F7 (Coagulation factor VII), 5-HT-1B (5-hydroxytryptamine receptor 1B), B-raf (Serine/threonine-protein kinase B-raf), and JAK2 (Tyrosine-protein kinase JAK2). Our experiment methodology follows \cite{lee2023exploring}. Specifically, we use the same ZINC 250k dataset \citep{irwin2005zinc}, which in our presented setup is used for fine-tuning with RL. For evaluation, we sample 3000 molecules and compute the same statistics, namely mean docking scores of top 5\% hit molecules and percentage of hit molecules. As done by \cite{lee2023exploring}, we define hit molecules as molecules with QED $>$ 0.5, SA $<$ 0.5 and docking scores less than the hit threshold of the specific target (-9.1, -10.3, -8.5, -10.0 and -8.78 kcal/mol for JAK2, B-raf, F7, PARP-1 and 5-HT-1B respectively). Our results are presented in Tables \ref{tab:mean_ds} and \ref{tab:percent_hits} alongside the best results reported from the MOOD model in \cite{lee2023exploring}. We refer to our model as \textbf{MoGeL}, which stands for \textbf{Mo}lecule \textbf{G}eneration with \textbf{L}atents. We fine-tuned the models Emb-128 and Emb-128 + P-64 (see section \ref{sec:model_scaling}). In this section, we will refer to these models as MoGeL-128 and MoGeL-64, with the numbers indicating the embedding dimension of the latent space, respectively. 
\begin{table}
\centering
    \caption{Mean Docking Scores (kcal/mol) and standard deviations (in brackets) of the top 5\% hit molecules from 3000 samples using different pre-trained and RL-fine-tuned models, along with the best results from the MOOD model. Lower docking scores correspond to stronger binding energies, indicating more stable protein-ligand complexes. The asterisk is used to indicate that, based on the standard deviation there is not much statistical difference with the best performing model.}
    \label{tab:mean_ds}
   \resizebox{\columnwidth}{!}{%
    \begin{tabular}{l|rrrrr}
    \midrule
    \multicolumn{1}{c|}{{\textbf{Model}}} & \multicolumn{1}{l|}{\textbf{JAK2}} & \multicolumn{1}{l|}{\textbf{F7}} & \multicolumn{1}{l|}{\textbf{PARP-1}} & \multicolumn{1}{l|}{\textbf{B-raf}} & \textbf{5-HT-1B} \\
    \midrule
    MOOD w/o OOD \citep{lee2023exploring} & \multicolumn{1}{l|}{-9.575 (0.075)} & \multicolumn{1}{l|}{-7.947 (0.034)} & \multicolumn{1}{l|}{-10.409 (0.030)} & \multicolumn{1}{l|}{-10.421 (0.05)} & -10.487 (0.068) \\
    \midrule
    MOOD \citep{lee2023exploring} & \multicolumn{1}{l|}{\textbf{-10.147 (0.060)}} & \multicolumn{1}{l|}{-8.160 (0.071)} & \multicolumn{1}{l|}{\textbf{-10.865 (0.113)}} & \multicolumn{1}{l|}{\textbf{-11.063 (0.034)}} & \textbf{-11.145 (0.042)} \\
    \midrule
    MoGeL-128 (ours) & \multicolumn{1}{l|}{-9.53 (0.34)} & \multicolumn{1}{l|}{-8.10 (0.24)} & \multicolumn{1}{l|}{-10.16 (0.36)} & \multicolumn{1}{l|}{-9.94 (0.34)} & -10.04 (0.32) \\
    \midrule
    MoGeL-128 + RL (ours) & \multicolumn{1}{l|}{-9.74 (0.31)} & \multicolumn{1}{l|}{-8.29 (0.29)} & \multicolumn{1}{l|}{-10.62 (0.33)} & \multicolumn{1}{l|}{-10.14 (0.49)} & -10.40 (0.39)\\
    \midrule
    MoGeL-64 (ours) & \multicolumn{1}{l|}{-9.71 (0.35)} & \multicolumn{1}{l|}{-8.33 (0.24)} & \multicolumn{1}{l|}{-10.49 (0.35)} & \multicolumn{1}{l|}{-10.52 (0.39)} & -10.40 (0.32) \\
    \midrule
    MoGeL-64 + RL (ours) & \multicolumn{1}{l|}{\textbf{-9.89 (0.41)*}} & \multicolumn{1}{l|}{\textbf{-8.59 (1.21)}} & \multicolumn{1}{l|}{\textbf{-10.72 (0.35)*}} & \multicolumn{1}{l|}{-10.57 (0.36)} & -10.52 (0.24) \\
    \midrule
    \end{tabular}%
    }
\end{table}

\begin{table}
 \centering
    \caption{Percentage of hits molecules (\%) from 3000 samples using different pre-trained and RL-fine-tuned models, along with the best results from the MOOD model}
    \label{tab:percent_hits}
    \resizebox{0.8\columnwidth}{!}{%
    \begin{tabular}{l|lllll}
    \midrule
    \multicolumn{1}{c|}{{\textbf{Model}}} & \multicolumn{1}{l|}{\textbf{JAK2}} & \multicolumn{1}{l|}{\textbf{F7}} & \multicolumn{1}{l|}{\textbf{PARP-1}} & \multicolumn{1}{l|}{\textbf{B-raf}} & \textbf{5-HT-1B} \\
    \midrule 
    MOOD w/o OOD \citep{lee2023exploring} & \multicolumn{1}{l|}{3.953} & \multicolumn{1}{l|}{0.433} & \multicolumn{1}{l|}{3.4} & \multicolumn{1}{l|}{2.207} & 11.873 \\
    \midrule
    MOOD \citep{lee2023exploring} & \multicolumn{1}{l|}{9.200} & \multicolumn{1}{l|}{0.733} & \multicolumn{1}{l|}{7.017} & \multicolumn{1}{l|}{\textbf{5.240}} & {\textbf{18.673}} \\
    \midrule
    MoGeL-128 (ours) & \multicolumn{1}{l|}{5.13} & \multicolumn{1}{l|}{0.27} & \multicolumn{1}{l|}{2.54} & \multicolumn{1}{l|}{0.46} & 1.89 \\
    \midrule
    MoGeL-128 + RL (ours) & \multicolumn{1}{l|}{8.91} & \multicolumn{1}{l|}{0.91} & \multicolumn{1}{l|}{\textbf{7.11}} & \multicolumn{1}{l|}{1.14} & 4.64 \\
    \midrule
    MoGeL-64 (ours) & \multicolumn{1}{l|}{7.35} & \multicolumn{1}{l|}{0.88} & \multicolumn{1}{l|}{5.84} & \multicolumn{1}{l|}{2.17} & 5.43 \\
    \midrule
    MoGeL-64 + RL (ours) & \multicolumn{1}{l|}{\textbf{11.03}} & \multicolumn{1}{l|}{\textbf{1.32}} & \multicolumn{1}{l|}{6.77} & \multicolumn{1}{l|}{3.48} & 7.86 \\
    \midrule
    \end{tabular}%
    }
\end{table}
Our results indicate that our pre-trained model performs on par with the MOOD w/o OOD model (i.e when MOOD is not trained to encourage out of training distribution sampling) on most of the targets except 5-HT-1B. Our RL fine-tuning improves the performance of our pre-trained model by almost doubling the percentage of hit molecules across all targets. The mean docking scores of our top 5\% of molecules closely match those of the MOOD model, and our models surpass MOOD in terms of the percentage of hits for the targets JAK2, F7, and PARP1-1.  This is an important result because our architecture allows for easy conditioning of the model based on various properties and exploration of different regions in the latent distribution.  
\begin{figure}
    \centering
    \includegraphics[width=0.9\linewidth]{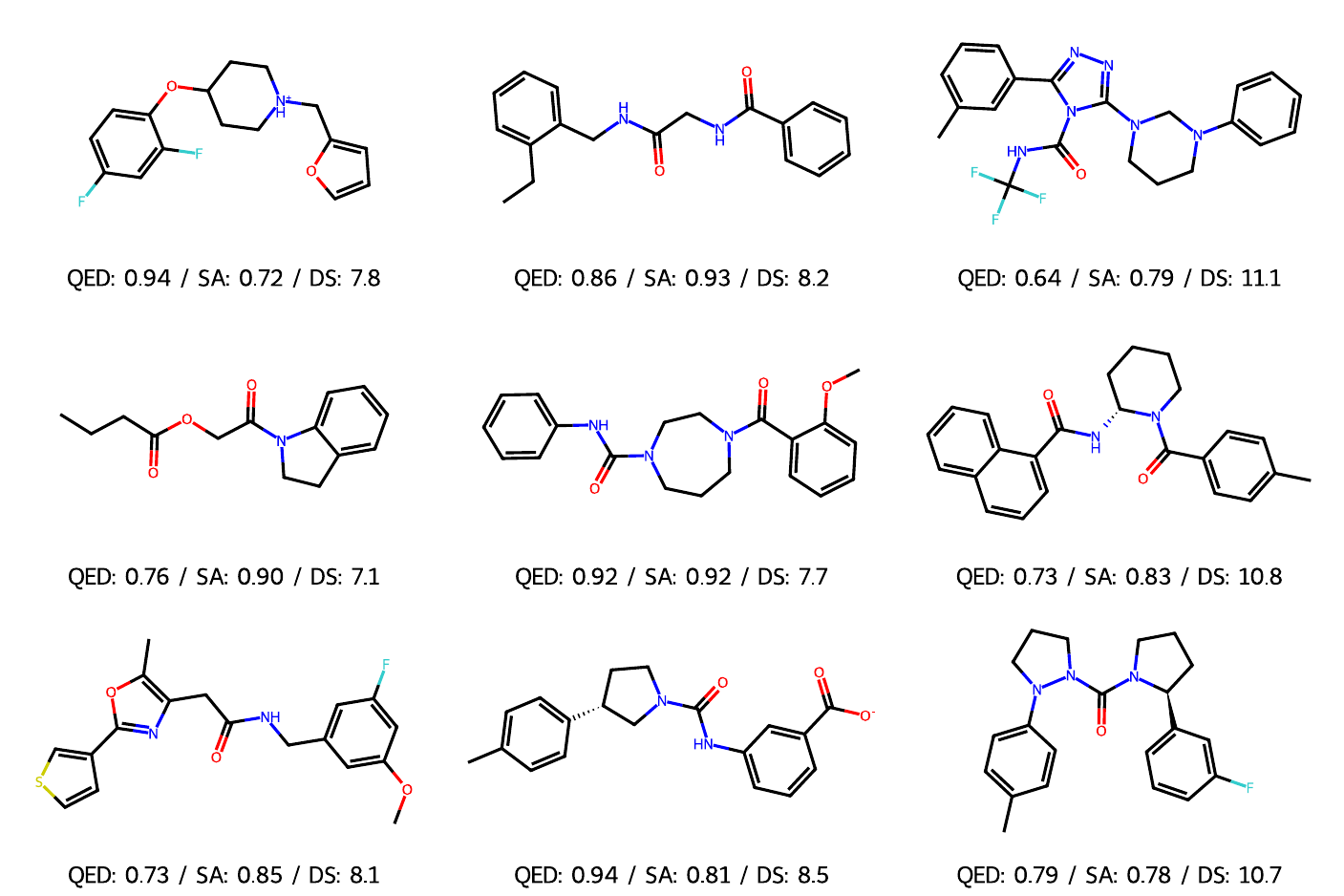}
    \caption{\textit{Visualisation of generated molecules.} \textbf{Left:} Original molecules from the dataset. \textbf{Middle:} Generated molecules with the pre-trained model. \textbf{Right:} The generated molecules obtained after RL fine-tuning. These results were obtained after fine-tuning MoGeL-64 for molecular docking to the target JAK2. The labels indicate the molecules measured QED, SA and docking score.}
    \label{fig:rl_mols}
\end{figure}
Fig. \ref{fig:rl_mols} displays some molecules from which we can generate hit molecules using the MoGeL-64 RL fine-tuned model to the target JAK2. These images validate our model's ability to generate molecules with QED values similar to those in our training set. We observed that the docking scores of the molecules before fine-tuning are very similar to the original docking scores, while there is a significant increase in docking scores after fine-tuning. Another interesting observation is that the original molecules and the fine-tuned molecules often contain similar types of functional groups and atoms but in different arrangements and numbers. The structures of the molecules are thus altered, with moieties added or removed to improve the docking scores, sometimes with major changes to the core structure --- the scaffold. This suggests that MoGeL is capable of both hit optimization and \textit{scaffold hopping}, i.e. improving around a scaffold or finding alternative structural motifs which retain or enhance target binding affinity \citep{bohm2004scaffoldhopping}.

\section{Discussion}
This work introduces a new generative model for small molecule generation, built on a transformer encoder-decoder architecture with a latent sampling space. We propose a molecular representation called SAFER, an adaptation of the SAFE representation, to train this model. Our experiments demonstrate the model's ability to generate small drug-like molecules and show that fine-tuning various parts of the model significantly improves generation of new molecules with high docking scores to specific protein targets. This flexibility is a key feature of our architecture. Future work will involve scaling the model, studying the latent distribution properties, and conditioning the fine-tuned model on protein target structural information. If this conditioning is successful, it could allow the model to generalise to generating molecules with high docking scores for unseen protein targets, improving the exploration of the chemical space in the drug discovery process.

\nocite{langley00}

\bibliography{example_paper}

\begin{thebibliography}{44}
\providecommand{\natexlab}[1]{#1}
\providecommand{\url}[1]{\texttt{#1}}
\expandafter\ifx\csname urlstyle\endcsname\relax
  \providecommand{\doi}[1]{doi: #1}\else
  \providecommand{\doi}{doi: \begingroup \urlstyle{rm}\Url}\fi

\bibitem[Alhossary et~al.(2015)Alhossary, Handoko, Mu, and Kwoh]{qvina2}
Alhossary, A., Handoko, S.~D., Mu, Y., and Kwoh, C.-K.
\newblock {Fast, accurate, and reliable molecular docking with QuickVina 2}.
\newblock \emph{Bioinformatics}, 31\penalty0 (13):\penalty0 2214--2216, 02 2015.
\newblock ISSN 1367-4803.
\newblock \doi{10.1093/bioinformatics/btv082}.
\newblock URL \url{https://doi.org/10.1093/bioinformatics/btv082}.

\bibitem[Atance et~al.(2022)Atance, Diez, Engkvist, Olsson, and Mercado]{atance2022novo}
Atance, S.~R., Diez, J.~V., Engkvist, O., Olsson, S., and Mercado, R.
\newblock De novo drug design using reinforcement learning with graph-based deep generative models.
\newblock \emph{Journal of Chemical Information and Modeling}, 62\penalty0 (20):\penalty0 4863--4872, 2022.

\bibitem[Bickerton et~al.(2012)Bickerton, Paolini, Besnard, Muresan, and Hopkins]{qed}
Bickerton, G.~R., Paolini, G.~V., Besnard, J., Muresan, S., and Hopkins, A.~L.
\newblock {{Q}uantifying the chemical beauty of drugs}.
\newblock \emph{Nat Chem}, 4\penalty0 (2):\penalty0 90--98, Jan 2012.

\bibitem[Bohacek et~al.(1996)Bohacek, McMartin, and Guida]{bohacek1996art}
Bohacek, R.~S., McMartin, C., and Guida, W.~C.
\newblock The art and practice of structure-based drug design: a molecular modeling perspective.
\newblock \emph{Medicinal research reviews}, 16\penalty0 (1):\penalty0 3--50, 1996.

\bibitem[Bongini et~al.(2021)Bongini, Bianchini, and Scarselli]{bongini2021molecular}
Bongini, P., Bianchini, M., and Scarselli, F.
\newblock Molecular generative graph neural networks for drug discovery.
\newblock \emph{Neurocomputing}, 450:\penalty0 242--252, 2021.

\bibitem[Bradbury et~al.(2018)Bradbury, Frostig, Hawkins, Johnson, Leary, Maclaurin, Necula, Paszke, Vander{P}las, Wanderman-{M}ilne, and Zhang]{jax2018github}
Bradbury, J., Frostig, R., Hawkins, P., Johnson, M.~J., Leary, C., Maclaurin, D., Necula, G., Paszke, A., Vander{P}las, J., Wanderman-{M}ilne, S., and Zhang, Q.
\newblock {JAX}: composable transformations of {P}ython+{N}um{P}y programs, 2018.
\newblock URL \url{http://github.com/google/jax}.

\bibitem[Böhm et~al.(2004)Böhm, Flohr, and Stahl]{bohm2004scaffoldhopping}
Böhm, H.-J., Flohr, A., and Stahl, M.
\newblock Scaffold hopping.
\newblock \emph{Drug Discovery Today: Technologies}, 1\penalty0 (3):\penalty0 217--224, 2004.
\newblock ISSN 1740-6749.
\newblock \doi{https://doi.org/10.1016/j.ddtec.2004.10.009}.
\newblock URL \url{https://www.sciencedirect.com/science/article/pii/S1740674904000460}.

\bibitem[Choi et~al.(2010)Choi, Cha, Tappert, et~al.]{choi2010survey}
Choi, S.-S., Cha, S.-H., Tappert, C.~C., et~al.
\newblock A survey of binary similarity and distance measures.
\newblock \emph{Journal of systemics, cybernetics and informatics}, 8\penalty0 (1):\penalty0 43--48, 2010.

\bibitem[Corso et~al.(2022)Corso, St{\"a}rk, Jing, Barzilay, and Jaakkola]{corso2022diffdock}
Corso, G., St{\"a}rk, H., Jing, B., Barzilay, R., and Jaakkola, T.
\newblock Diffdock: Diffusion steps, twists, and turns for molecular docking.
\newblock \emph{arXiv preprint arXiv:2210.01776}, 2022.

\bibitem[Degen et~al.(2008)Degen, Wegscheid-Gerlach, Zaliani, and Rarey]{degen2008brics}
Degen, J., Wegscheid-Gerlach, C., Zaliani, A., and Rarey, M.
\newblock On the art of compiling and using 'drug-like' chemical fragment spaces.
\newblock \emph{ChemMedChem}, 3\penalty0 (10):\penalty0 1503--1507, 2008.
\newblock \doi{https://doi.org/10.1002/cmdc.200800178}.
\newblock URL \url{https://chemistry-europe.onlinelibrary.wiley.com/doi/abs/10.1002/cmdc.200800178}.

\bibitem[Du et~al.(2022)Du, Fu, Sun, and Liu]{du2022molgensurvey}
Du, Y., Fu, T., Sun, J., and Liu, S.
\newblock Molgensurvey: A systematic survey in machine learning models for molecule design.
\newblock \emph{arXiv preprint arXiv:2203.14500}, 2022.

\bibitem[Fang et~al.(2021)Fang, Zeng, Liu, Bo, Dong, and Chen]{fang2021transformer}
Fang, L., Zeng, T., Liu, C., Bo, L., Dong, W., and Chen, C.
\newblock Transformer-based conditional variational autoencoder for controllable story generation.
\newblock \emph{arXiv preprint arXiv:2101.00828}, 2021.

\bibitem[Feng et~al.(2024)Feng, Zhang, Deng, and Xiong]{feng2024gcardti}
Feng, Y., Zhang, Y., Deng, Z., and Xiong, M.
\newblock Gcardti: Drug--target interaction prediction based on a hybrid mechanism in drug selfies.
\newblock \emph{Quantitative Biology}, 2024.

\bibitem[Fromer \& Coley(2023)Fromer and Coley]{FROMER2023100678}
Fromer, J.~C. and Coley, C.~W.
\newblock Computer-aided multi-objective optimization in small molecule discovery.
\newblock \emph{Patterns}, 4\penalty0 (2):\penalty0 100678, 2023.
\newblock ISSN 2666-3899.
\newblock \doi{https://doi.org/10.1016/j.patter.2023.100678}.
\newblock URL \url{https://www.sciencedirect.com/science/article/pii/S2666389923000016}.

\bibitem[Ganea et~al.(2021)Ganea, Huang, Bunne, Bian, Barzilay, Jaakkola, and Krause]{ganea2021independent}
Ganea, O.-E., Huang, X., Bunne, C., Bian, Y., Barzilay, R., Jaakkola, T., and Krause, A.
\newblock Independent {SE}(3)-equivariant models for end-to-end rigid protein docking.
\newblock \emph{arXiv preprint arXiv:2111.07786}, 2021.

\bibitem[G{\'o}mez-Bombarelli et~al.(2018)G{\'o}mez-Bombarelli, Wei, Duvenaud, Hern{\'a}ndez-Lobato, S{\'a}nchez-Lengeling, Sheberla, Aguilera-Iparraguirre, Hirzel, Adams, and Aspuru-Guzik]{gomez2018automatic}
G{\'o}mez-Bombarelli, R., Wei, J.~N., Duvenaud, D., Hern{\'a}ndez-Lobato, J.~M., S{\'a}nchez-Lengeling, B., Sheberla, D., Aguilera-Iparraguirre, J., Hirzel, T.~D., Adams, R.~P., and Aspuru-Guzik, A.
\newblock Automatic chemical design using a data-driven continuous representation of molecules.
\newblock \emph{ACS central science}, 4\penalty0 (2):\penalty0 268--276, 2018.

\bibitem[Gupta et~al.(2018)Gupta, M{\"u}ller, Huisman, Fuchs, Schneider, and Schneider]{gupta2018generative}
Gupta, A., M{\"u}ller, A.~T., Huisman, B.~J., Fuchs, J.~A., Schneider, P., and Schneider, G.
\newblock Generative recurrent networks for de novo drug design.
\newblock \emph{Molecular informatics}, 37\penalty0 (1-2):\penalty0 1700111, 2018.

\bibitem[Irwin \& Shoichet(2005)Irwin and Shoichet]{irwin2005zinc}
Irwin, J.~J. and Shoichet, B.~K.
\newblock {ZINC}- a free database of commercially available compounds for virtual screening.
\newblock \emph{Journal of chemical information and modeling}, 45\penalty0 (1):\penalty0 177--182, 2005.

\bibitem[Irwin et~al.(2022)Irwin, Dimitriadis, He, and Bjerrum]{irwin2022chemformer}
Irwin, R., Dimitriadis, S., He, J., and Bjerrum, E.~J.
\newblock Chemformer: a pre-trained transformer for computational chemistry.
\newblock \emph{Machine Learning: Science and Technology}, 3\penalty0 (1):\penalty0 015022, 2022.

\bibitem[Jaegle et~al.(2021)Jaegle, Borgeaud, Alayrac, Doersch, Ionescu, Ding, Koppula, Zoran, Brock, Shelhamer, et~al.]{jaegle2021perceiver}
Jaegle, A., Borgeaud, S., Alayrac, J.-B., Doersch, C., Ionescu, C., Ding, D., Koppula, S., Zoran, D., Brock, A., Shelhamer, E., et~al.
\newblock Perceiver io: A general architecture for structured inputs \& outputs.
\newblock \emph{arXiv preprint arXiv:2107.14795}, 2021.

\bibitem[Jeon \& Kim(2020)Jeon and Kim]{jeon2020autonomous}
Jeon, W. and Kim, D.
\newblock Autonomous molecule generation using reinforcement learning and docking to develop potential novel inhibitors.
\newblock \emph{Scientific reports}, 10\penalty0 (1):\penalty0 22104, 2020.

\bibitem[Jiang et~al.(2021)Jiang, Wu, Hsieh, Chen, Liao, Wang, Shen, Cao, Wu, and Hou]{jiang2021could}
Jiang, D., Wu, Z., Hsieh, C.-Y., Chen, G., Liao, B., Wang, Z., Shen, C., Cao, D., Wu, J., and Hou, T.
\newblock Could graph neural networks learn better molecular representation for drug discovery? a comparison study of descriptor-based and graph-based models.
\newblock \emph{Journal of cheminformatics}, 13:\penalty0 1--23, 2021.

\bibitem[Jo et~al.(2022)Jo, Lee, and Hwang]{jo2022score}
Jo, J., Lee, S., and Hwang, S.~J.
\newblock Score-based generative modeling of graphs via the system of stochastic differential equations.
\newblock In \emph{International Conference on Machine Learning}, pp.\  10362--10383. PMLR, 2022.

\bibitem[Kingma \& Ba(2014)Kingma and Ba]{adam}
Kingma, D.~P. and Ba, J.
\newblock Adam: A method for stochastic optimization.
\newblock \emph{arXiv preprint arXiv:1412.6980}, 2014.

\bibitem[Krenn et~al.(2020)Krenn, H{\"a}se, Nigam, Friederich, and Aspuru-Guzik]{krenn2020self}
Krenn, M., H{\"a}se, F., Nigam, A., Friederich, P., and Aspuru-Guzik, A.
\newblock Self-referencing embedded strings ({SELFIES}): A 100\% robust molecular string representation.
\newblock \emph{Machine Learning: Science and Technology}, 1\penalty0 (4):\penalty0 045024, 2020.

\bibitem[Landrum et~al.(2013)]{landrum2013rdkit}
Landrum, G. et~al.
\newblock Rdkit: A software suite for cheminformatics, computational chemistry, and predictive modeling.
\newblock \emph{Greg Landrum}, 8\penalty0 (31.10):\penalty0 5281, 2013.

\bibitem[Langley(2000)]{langley00}
Langley, P.
\newblock Crafting papers on machine learning.
\newblock In Langley, P. (ed.), \emph{Proceedings of the 17th International Conference on Machine Learning (ICML 2000)}, pp.\  1207--1216, Stanford, CA, 2000. Morgan Kaufmann.

\bibitem[Lee et~al.(2023)Lee, Jo, and Hwang]{lee2023exploring}
Lee, S., Jo, J., and Hwang, S.~J.
\newblock Exploring chemical space with score-based out-of-distribution generation.
\newblock In \emph{International Conference on Machine Learning}, pp.\  18872--18892. PMLR, 2023.

\bibitem[Luo et~al.(2021)Luo, Yan, and Ji]{luo2021graphdf}
Luo, Y., Yan, K., and Ji, S.
\newblock Graphdf: A discrete flow model for molecular graph generation.
\newblock In \emph{International conference on machine learning}, pp.\  7192--7203. PMLR, 2021.

\bibitem[Madhawa et~al.(2019)Madhawa, Ishiguro, Nakago, and Abe]{madhawa2019graphnvp}
Madhawa, K., Ishiguro, K., Nakago, K., and Abe, M.
\newblock Graphnvp: An invertible flow model for generating molecular graphs.
\newblock \emph{arXiv preprint arXiv:1905.11600}, 2019.

\bibitem[Noutahi et~al.(2024)Noutahi, Gabellini, Craig, Lim, and Tossou]{noutahi2024gotta}
Noutahi, E., Gabellini, C., Craig, M., Lim, J.~S., and Tossou, P.
\newblock Gotta be safe: A new framework for molecular design.
\newblock \emph{Digital Discovery}, 2024.

\bibitem[O'Boyle \& Dalke(2018)O'Boyle and Dalke]{o2018deepsmiles}
O'Boyle, N. and Dalke, A.
\newblock Deepsmiles: An adaptation of {SMILES} for use in machine-learning of chemical structures.
\newblock \emph{ChemRxiv}, 2018.
\newblock \doi{10.26434/chemrxiv.7097960.v1}.

\bibitem[Olivecrona et~al.(2017)Olivecrona, Blaschke, Engkvist, and Chen]{olivecrona2017molecular}
Olivecrona, M., Blaschke, T., Engkvist, O., and Chen, H.
\newblock Molecular de-novo design through deep reinforcement learning.
\newblock \emph{Journal of cheminformatics}, 9:\penalty0 1--14, 2017.

\bibitem[Podda et~al.(2020)Podda, Bacciu, and Micheli]{podda2020deep}
Podda, M., Bacciu, D., and Micheli, A.
\newblock A deep generative model for fragment-based molecule generation.
\newblock In \emph{International conference on artificial intelligence and statistics}, pp.\  2240--2250. PMLR, 2020.

\bibitem[Polishchuk et~al.(2013)Polishchuk, Madzhidov, and Varnek]{polishchuk2013estimation}
Polishchuk, P.~G., Madzhidov, T.~I., and Varnek, A.
\newblock Estimation of the size of drug-like chemical space based on gdb-17 data.
\newblock \emph{Journal of computer-aided molecular design}, 27:\penalty0 675--679, 2013.

\bibitem[Powers et~al.(2022)Powers, Yu, Suriana, and Dror]{powers2022fragment}
Powers, A.~S., Yu, H.~H., Suriana, P., and Dror, R.~O.
\newblock Fragment-based ligand generation guided by geometric deep learning on protein-ligand structure.
\newblock \emph{bioRxiv}, pp.\  2022--03, 2022.

\bibitem[Satorras et~al.(2021)Satorras, Hoogeboom, and Welling]{satorras2021n}
Satorras, V.~G., Hoogeboom, E., and Welling, M.
\newblock E(n) equivariant graph neural networks.
\newblock In \emph{International conference on machine learning}, pp.\  9323--9332. PMLR, 2021.

\bibitem[Skinnider(2024)]{skinnider2024invalid}
Skinnider, M.~A.
\newblock Invalid smiles are beneficial rather than detrimental to chemical language models.
\newblock \emph{Nature Machine Intelligence}, 6\penalty0 (4):\penalty0 437--448, 2024.

\bibitem[Tarasov et~al.(2023)Tarasov, Mbou~Sob, Arbesu, Siboni, Boyer, Skwark, Smit, Bent, and Pretorius]{tarasov2023offline}
Tarasov, D., Mbou~Sob, U.~A., Arbesu, M., Siboni, N., Boyer, S., Skwark, M., Smit, A., Bent, O., and Pretorius, A.
\newblock Offline {RL} for generative design of protein binders.
\newblock \emph{bioRxiv}, pp.\  2023--11, 2023.

\bibitem[Trott \& Olson(2010)Trott and Olson]{trott2010autodock}
Trott, O. and Olson, A.~J.
\newblock Autodock {V}ina: improving the speed and accuracy of docking with a new scoring function, efficient optimization, and multithreading.
\newblock \emph{Journal of computational chemistry}, 31\penalty0 (2):\penalty0 455--461, 2010.

\bibitem[Vaswani et~al.(2017)Vaswani, Shazeer, Parmar, Uszkoreit, Jones, Gomez, Kaiser, and Polosukhin]{vaswani2017attention}
Vaswani, A., Shazeer, N., Parmar, N., Uszkoreit, J., Jones, L., Gomez, A.~N., Kaiser, {\L}., and Polosukhin, I.
\newblock Attention is all you need.
\newblock \emph{Advances in neural information processing systems}, 30, 2017.

\bibitem[Weininger(1988)]{smiles}
Weininger, D.
\newblock {SMILES}, a chemical language and information system. 1. introduction to methodology and encoding rules.
\newblock \emph{Journal of Chemical Information and Computer Sciences}, 28\penalty0 (1):\penalty0 31--36, 1988.
\newblock \doi{10.1021/ci00057a005}.
\newblock URL \url{https://doi.org/10.1021/ci00057a005}.

\bibitem[Williams(1992)]{williams1992simple}
Williams, R.~J.
\newblock Simple statistical gradient-following algorithms for connectionist reinforcement learning.
\newblock \emph{Machine learning}, 8:\penalty0 229--256, 1992.

\bibitem[Zang \& Wang(2020)Zang and Wang]{zang2020moflow}
Zang, C. and Wang, F.
\newblock Moflow: an invertible flow model for generating molecular graphs.
\newblock In \emph{Proceedings of the 26th ACM SIGKDD international conference on knowledge discovery \& data mining}, pp.\  617--626, 2020.

\end{thebibliography}
\bibliographystyle{icml2024}

\newpage
\appendix
\onecolumn
\section{Appendix}
\subsection{Conversion Algorithm}
\begin{algorithm}[H]
\caption{Conversion from SMILES to SAFER representation}\label{smiles_to_safe}
\begin{algorithmic}[1]
\Procedure{ConvertSAFE}{\textit{molecule}}
    \State {\color{blue} \textit{molecule} $\gets$ Standardized \textit{molecule}}
    \State \textit{ring\_digits} $\gets$ extract all unique ring digits from \textit{molecule}
    \State \textit{fragments} $\gets$ fragment \textit{molecule} on specified bonds \Comment{We use BRICS bonds here}
    \State \textit{fragments\_str} $\gets$ $\{\}$
    \State \sout{Sort \textit{fragments} by size in descending order}
    \For{\textit{each frag} in \textit{fragments}}
        \State Add SMILES of \textit{frag} to \textit{fragments\_str}
    \EndFor
    \State \textit{safe\_str} $\gets$ join all elements in \textit{fragments\_str} with "."
    \State \textit{attach\_pos} $\gets$ extract all attachment points from \textit{safe\_str}
    \State \sout{\textit{i} $\gets$ max(\textit{ring\_digits}) + 1 \Comment{Find the next possible ring digits}}
    \For{\textit{each attach} in \textit{attach\_pos}}
        \State \sout{Replace \textit{attach} in \textit{safe\_str} with \textit{i}}
        \State \sout{\textit{i} $\gets$ \textit{i} + 1}
        \State {\color{blue} Replace \textit{opening attachment} in \textit{safe\_str} with "[@]"}
        \State {\color{blue} Replace \textit{closing attachment} in \textit{safe\_str} with "[@@]"}
    \EndFor
    \State \Return \textit{safe\_str}
\EndProcedure
\end{algorithmic}
\end{algorithm}

\subsection{Datasets}\label{model_impl}
For model pre-training, we use a large publicly available SMILES dataset. Specifically, we make use of the SAFE dataset on HuggingFace (\cite{noutahi2024gotta}). This dataset is one of the largest publicly available SMILES datasets, with over 1 billion molecules, curated from various datasets such as the ZINC and UniCHEM libraries \citep{noutahi2024gotta}. It consists of various molecular types ranging from drug-like molecules, peptides, multi-fragment molecules, polymers, reagents and non-small molecules. 

In our pre-training pipeline, we select a subset of the data consisting of the first 21 parquet files. Since the data is provided in the SAFE representation, we use the SAFE package\footnote{SAFE-github: \url{https://github.com/datamol-io/safe/}} to convert the molecules to their corresponding SMILES representation, and then convert to our SAFER representation. We discard any molecule that fails the SAFE to SMILES conversion and we filter all molecules to have a maximum of 140 tokens. This results in the final dataset consisting of roughly 150 million small molecules.

In order to fine-tune the models, we used the zinc250k dataset \citep{irwin2005zinc}. This dataset comprises 250k small drug-like molecules from the ZINC database. While the SAFE dataset we used for pre-training already includes molecules from the ZINC database, we specifically utilized the zinc250k dataset for fine-tuning to maintain consistency with the MOOD model \cite{lee2023exploring}, which we are using for comparison. We created a SAFER version of this dataset for this purpose.

\subsection{Model's implementation and pre-training details}\label{sl_training_details}
To train the entire network we use the following loss function:
\begin{equation}
   \mathcal{L} =  \sum^{|\mathcal{D}|}_{x_{1:L} \in \mathcal{D}}\sum_{i=1}^{L} \log p(x_i | x_{1:i-1}, z_{1:L}) \, - \, \text{KL}\left(\mathcal{N}(\mu_{1:L}, \text{diag}(\sigma^2_{1:L})) \,\middle\|\, \mathcal{N}(0, I)\right), \label{loss_eq}
\end{equation}
where $x_{1:L}$ is a molecule sequence of length $L$ sampled from the dataset $\mathcal{D}$ and $z_{1:L}$ is the corresponding latent vector sampled from a multivariate Gaussian distribution $\mathcal{N}(\mu_{1:L}, \text{diag}(\sigma^2_{1:L}))$ and used during cross attention. The loss function consists of two main terms. The first term is the canonical self-supervised language model loss, which in our setting can also be interpreted as a type of reconstruction loss which measures the model's ability to reconstruct the input sequence. The second term is a regularisation term that aims to maintain the learned latent space distribution as close as possible to a prior distribution which in this case is a standard Gaussian distribution. Hence to generate molecules we sample random embeddings from a Gaussian distribution and pass that to the decoder network to autoregressively generate a new molecule.

We implement and train our model using JAX \citep{jax2018github} taking advantage of JAX accelerator functions. We use the pre-trained SAFE BPE tokenizer\footnote{Tokenizer: \url{https://huggingface.co/datamol-io/safe-gpt}}. This tokenizer is adapted to include the two additional tokens required for the SAFER representation. We used 95 \% of the data from training and 5 \% for validation.
The models were trained on a v3-8 Google TPU. All model training was performed in parallel across each of the 8 TPU devices using a batch size of 128 per device with a mini-batch size of 16 (resulting in gradient accumulation of 8). Both the encoder and decoder are N-layer transformer models with learned positional embeddings. The decoder output layer is supplemented with a linear head. The encoder's output layer is linked to either the perceiver layer or the \texttt{mean} and \texttt{logvar} layers. The \texttt{mean} and \texttt{logvar} layers are implemented as single linear layers. Table \ref{tab:sl_hyp} presents the configuration details and hyperparameters for each of our pre-trained models.
\begin{table}[!ht]
\centering
\caption{Model parameters}\label{tab:sl_hyp}
\begin{tabular}{cll}
\toprule
\multirow{3}{*}{Shared parameters}  & Encoder layers  & 8\\
                                & Decoder layers &  8 \\
                                & Attention heads &  8 \\ 
\midrule
\multirow{3}{*}{Emb-128 / Emb-128 + P-64}  &  Embedding dim  & 128\\
                                & Forward feedward dim & 512 \\
                                & Final dim &  128 / 64 \\
                                & Perceiver layer & False / True \\
\midrule
\multirow{3}{*}{Emb-256 / Emb-256 + P-128}  &  Embedding dim  & 256\\
                                & Forward feedward dim & 1024 \\
                                & Final dim &  256 / 128 \\
                                & Perceiver layer & False / True \\
\midrule
\multirow{3}{*}{Hyperparameters}  & Batch size   & 128       \\
                                    & Gradient accumulations & 8 \\
                                      & Optimizer & Adam~\citep{adam} \\
                                      & Learning rate   & 5e-5      \\
                                      & Training epochs & 1 \\                                                   
\bottomrule
\end{tabular}
\vspace{6pt}
\end{table}

\subsection{Evalution metrics}\label{sec:eval_metrics}
\begin{enumerate}
\item \textbf{Validity rate}: The percentage of syntactically valid SMILES i.e.\ the number of molecule sequences we are able to successfully construct from their SMILES representation using software such as RDKit \citep {landrum2013rdkit}.
\item \textbf{Fragmentation rate}: The number of molecules with SMILES representations that lead to fragmented (disconnected) molecules.
\item \textbf{Uniqueness}: The percentage of unique SMILES representations in a set of molecules.
\item \textbf{Similarity}: To measure the novelty of the generated molecule, we use RDKit to compute the Tanimoto similarity distance between the Morgan fingerprints of generated samples and the molecules in the training set (see \citet{choi2010survey}). Due to the large size of the training set, we only compare against a subset of the training set molecules, hence we do not put too much emphasis on this metric in our analysis. 
\item \textbf{Quantitative Estimate of Drug-likeness (QED)}: The likelihood of a generated molecule to have molecular properties similar to known drugs \citep{qed}.
\item \textbf{Synthetic Accessibility (SA)}: A metric that scores the difficulty of experimentally synthesising the generated molecule from its fragments.
\end{enumerate}

\subsection{SAFE vs SAFER}
This experiment compares the SAFE tokenizer with our SAFER tokenizer when used to train our model. For this experiment, we choose an embedding dimension of 128 and we do not include the perceiver layer. Each model is trained for one epoch using the reduced SAFE dataset.

\begin{figure}[ht]
    \centering
    \subfigure[Greedy decoding]{%
        \label{greedy}%
        \includegraphics[width=0.45\textwidth]{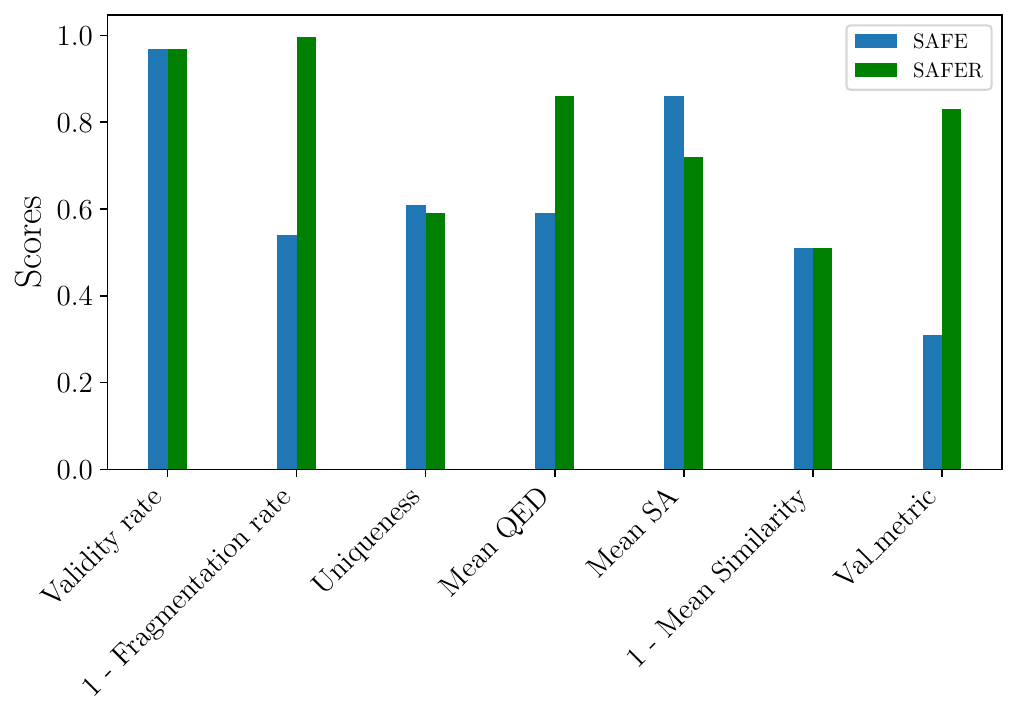}}%
    \hfill
    \subfigure[Stochastic decoding]{%
        \label{stochastic}%
        \includegraphics[width=0.45\textwidth]{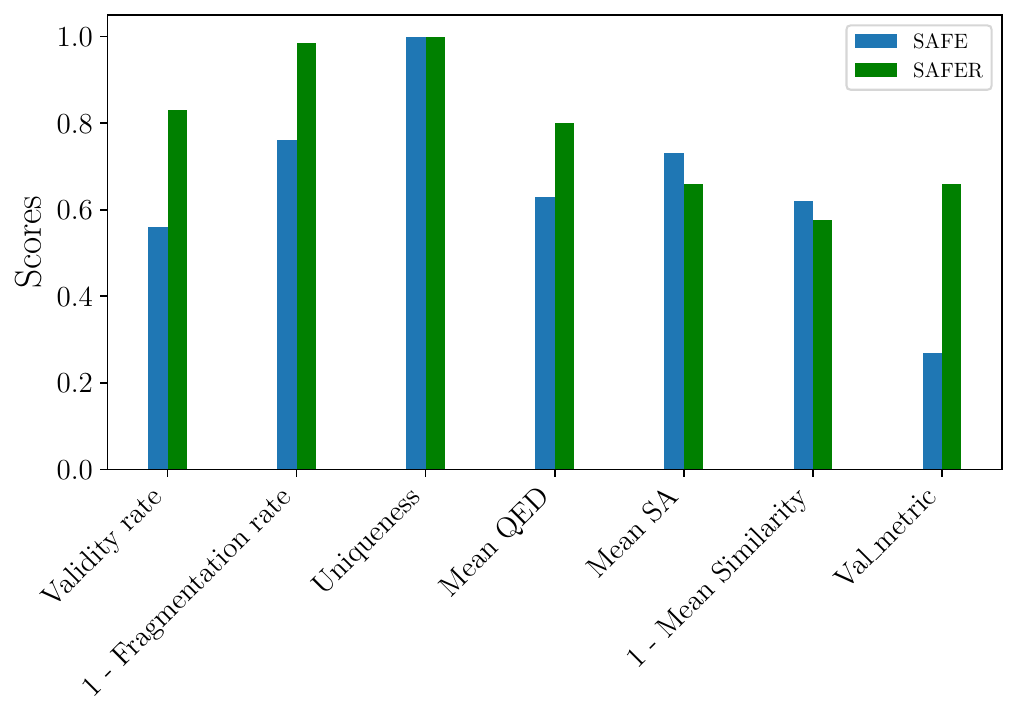}}%
    \caption{\textit{Comparing SAFE vs SAFER}. We compute the average scores of 10k molecules sampled using two models with embedding dimension 128 trained using the SAFE and SAFER representations. \textbf{Left:} Greedy decoding (temperature = 0). \textbf{Right:} Stochastic decoding (temperature = 1). Both representations are capable of generating molecules with high QED $>$ 0.5 but the SAFER representation significantly outperforms the SAFE representation on our combined validation metric (see Eq. \ref{sl_metric}) due to the lower fraction of fragmented molecules that are generated using the SAFER representation.}
    \label{fig:safe_modified_safe}
\end{figure}

Our results are shown in Figure \ref{fig:safe_modified_safe} comparing SAFE and SAFER representations across various metrics. Both representations are capable of generating high-quality molecules. However, our SAFER representation tends to produce more chemically plausible molecules when trained with our architecture. In contrast, the SAFE representation exhibits a fragmentation rate of almost 40\% with greedy decoding and a validity rate of less than 60\% with stochastic decoding (temperature = 1) compared to a fragmentation rate of less than 1\% with greedy decoding and a validity rate greater than 80\% with stochastic decoding for our SAFER representation.

\subsection{Reinforcement Learning (RL) fine-tuning}
\begin{figure}[H]
    \centering
    \includegraphics[width=1\linewidth]{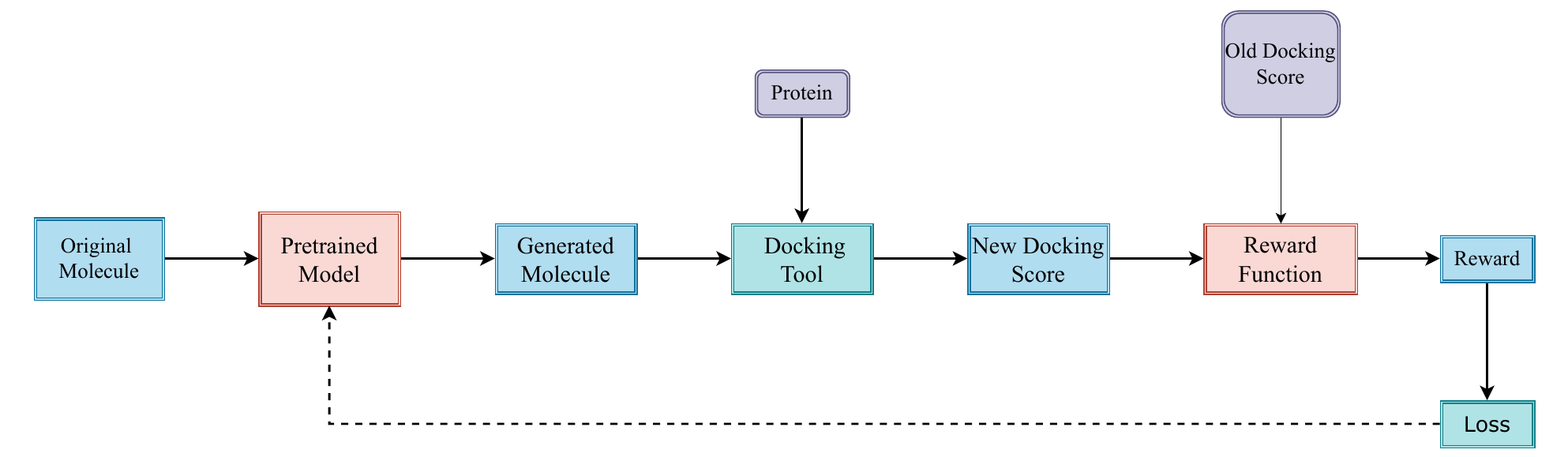}
    \caption{Schematic representation of the RL fine-tuning pipeline. The original molecule is passed to the pre-trained model that maps it to a region in our latent distribution using the \texttt{mean} and \texttt{logvar} layers. A latent vector is sampled for this region and passed to the decoder to generate a new molecule. The new molecule and the protein target are passed to the docking tool to perform molecular docking and produce a docking score for the new molecule. Following this, a reward is assigned to the molecule based on the comparison between the new docking score and the original docking score. We then use the reward to compute the loss and update the model's parameters.}
    \label{fig:Rl Finetune}
\end{figure}
For RL fine-tuning, we start by initializing both the sampling policy and the training policy with the best pre-trained checkpoint that we want to fine-tune. The policy model is used to generate samples from the input molecules in our dataset. The data loader is designed such that each input molecule is utilized multiple times to generate multiple experiences for each molecule. We use the Python wrapper tool Qvina2 \citep{qvina2} to dock the generated molecules to the target proteins and compute the docking scores. Qvina2's multi-processing capability allows us to dock multiple molecules at once. All molecules with failed docking attempts are assigned a dummy docking score of -99, which is used for masking during the loss computation. We fine-tuned the model with various learning rates and found that either 5e-4 or 5e-5 performed the best for each of the targets. Every 10 training iterations, the sampling policy is updated to the training policy. We run different experiments and fine-tuned each model with multiple seeds to vary that model doesn't perform well only for specific seeds.

\begin{figure}%
    \centering
    \subfigure[Mean docking scores]{%
        \label{mean_docking_scores}%
        \includegraphics[width=0.45\textwidth]{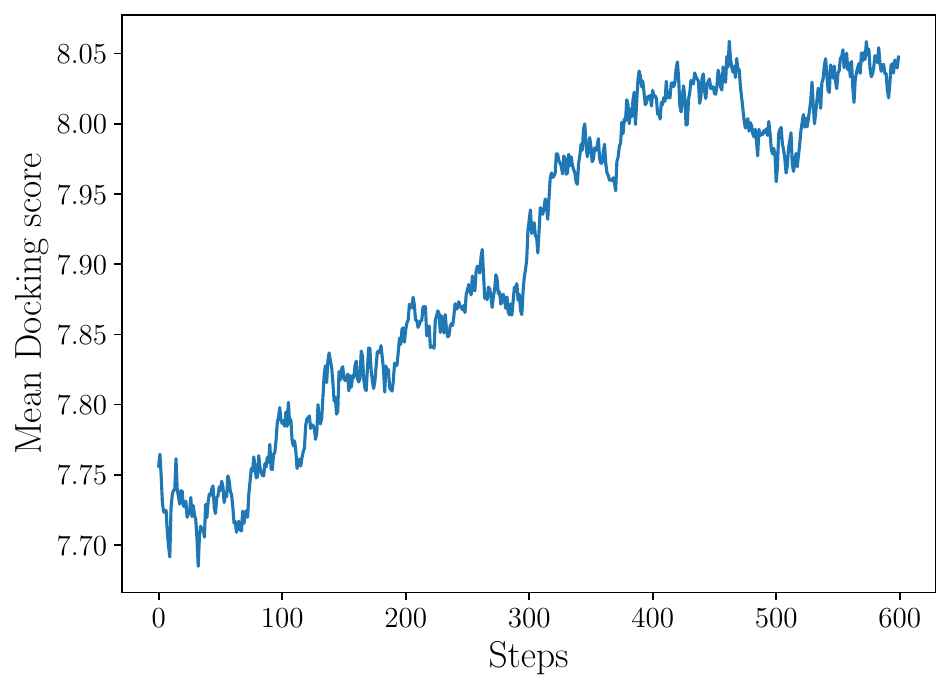}}%
    \hfill
    \subfigure[Molecules with docking scores change $>$ 2]{%
        \label{increase}%
        \includegraphics[width=0.45\textwidth]{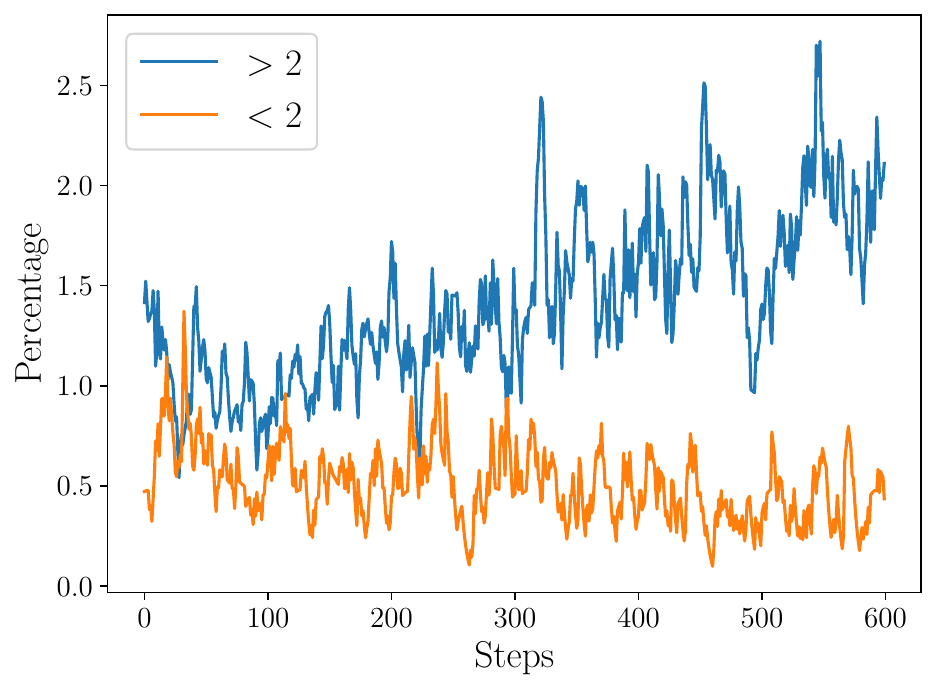}}%
    \caption{(a) Mean docking scores on the validation set during the fine-tuning of the model MoGel-128 to the protein target JAK2. The blue and orange curves in (b) are plots of the percentage of molecules whose docking scores increase or reduce by values greater than 2, respectively.}
    \label{fig:jak2_rl_metrics}
\end{figure}

Fig. \ref{fig:jak2_rl_metrics} show some training metrics for the fine-tuning of the model MoGel-128 to the target JAK2. These metrics were obtained using samples from our validation dataset and show a steady increase in the mean docking score of the samples. Additionally, from Fig. \ref{increase}, we can observe that the fraction of molecules whose docking scores decrease by 2 does not increase during training, unlike the fraction of molecules whose docking scores increase by 2. This suggests that our reward model is capable of guiding our model's latent space to sample molecules with higher docking scores.

\begin{figure}
    \centering
    \includegraphics[width=0.9\linewidth]{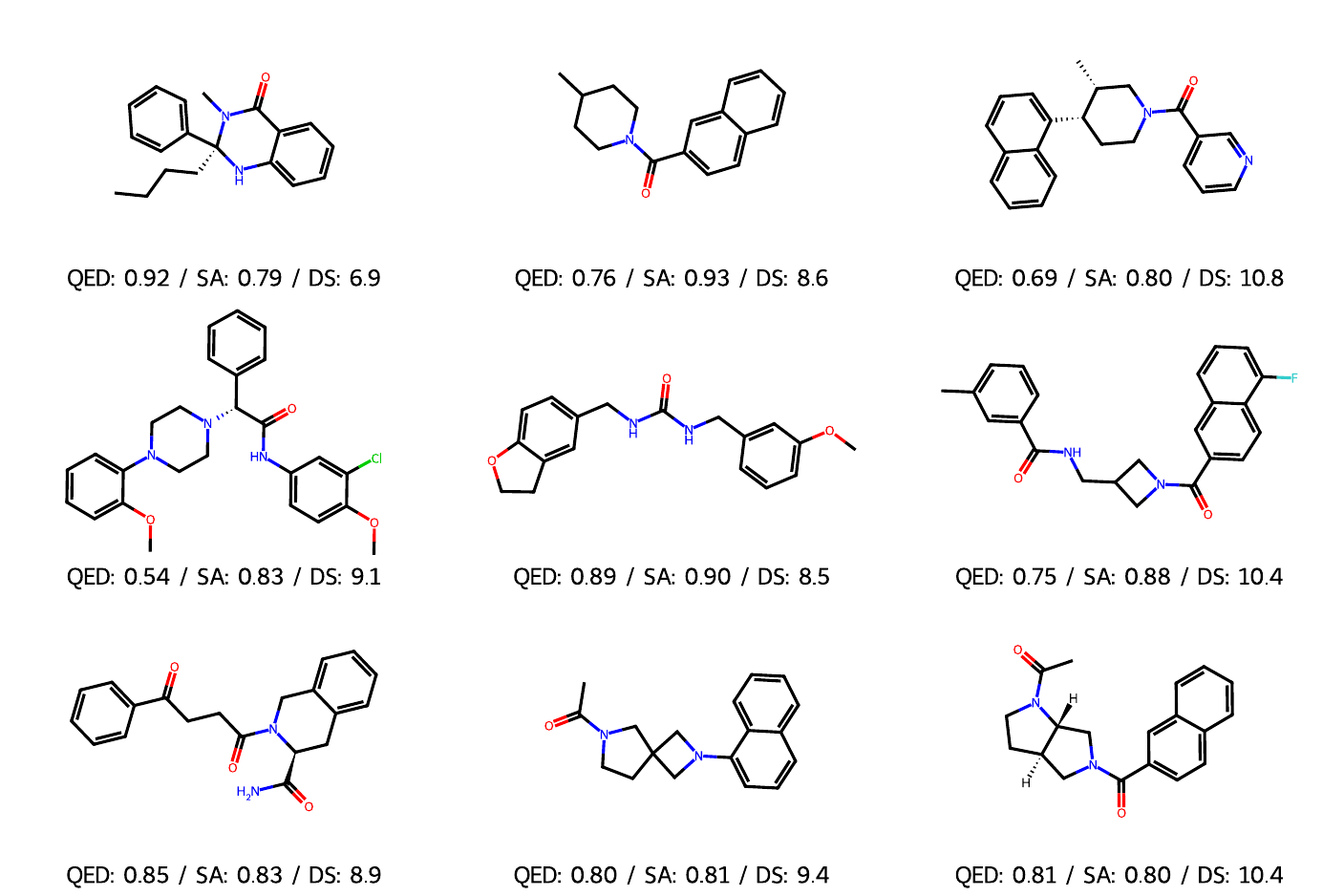}
    \caption{\textit{Visualisation of generated molecules.} \textbf{Left:} Original molecules from the dataset. \textbf{Middle:} generated molecules with the pre-trained model. \textbf{Right:} The generated molecules obtained after RL fine-tuning. These results were obtained after fine-tuning MoGel-128 for molecular docking to the target JAK2. The labels indicate the molecules measured QED, SA and docking score.}
    \label{fig:rl_mols2}
\end{figure}

Fig. \ref{fig:rl_mols2} displays example hit molecules obtained after MoGel-128 is fine-tuned to the JAK2 target alongside the original molecules, the pre-trained generated molecule, their QED, SA, and docking scores.


\end{document}